\input phyzzx.tex
\tolerance=1000
\voffset=-0.0cm
\hoffset=0.7cm
\sequentialequations
\def\rl{\rightline}

\def\t1{{\tilde 1}}

\def\SUSY{supersymmetry }

\def\t{\theta}

\REF{\GRA}{A. H. Chamseddine, R. Arnowitt and P. Nath, Phys. Rev. Lett. {\bf 49} (1982) 970; R. Barbieri, S. Ferrara and C. A. Savoy,
Phys. Lett. {\bf B119} (1982) 343; L. E. Ibanez, Phys. Lett. {\bf B118} (1982) 73; L. J. Hall, J. D. Lykken and S. Weinberg, Phys. Rev. {\bf D27} (1983) 2359; N. Ohta, Prog. Theor. Phys. {\bf 70} (1983) 542; L. Alvarez-Gaume, J. Polchinski and M. Wise, Nucl. Phys.
{\bf B221} (1983) 495.}
\REF{\OGM}{M. Dine, W. Fischler and M. Srednicki, Nucl. Phys. {\bf B189} (1981) 575; S. Dimopoulos and S. Raby, Nucl. Phys. {\bf B192} (1982) 353; M. Dine and W. Fischler, Phys. Lett. {\bf B110} (1982) 227; Nucl. Phys. {\bf B204} (1982) 346; C. Nappi and B. Ovrut, Phys. Lett. {\bf B113} (1982) 175; L. Alvarez-Gaume,
M. Claudson and M. Wise, Nucl. Phys. {\bf B207} (1982) 96; S. Dimopoulos and S. Raby, Nucl. Phys. {\bf B219} (1983) 479.}
\REF{\GM}{M. Dine and A. E. Nelson, Phys. Rev. {\bf D48} (1993) 1277, [arXiv:hep-ph/9303230]; M. Dine, A. E. Nelson and Y. Shirman, Phys. Rev. {\bf D51} (1995) 1362, 
[arXiv:hep-ph/9408384]; M. Dine, A. E. Nelson, Y. Nir and Y. Shirman, Phys. Rev. {\bf D53} (1996) 2658, [arXiv:hep-ph/9507378].} 
\REF{\GUI}{G. F. Guidice and R. Rattazzi, Phys. Rept. {\bf 322} (1999) 419, [arXiv:hep-ph/9801271].}
\REF{\AM}{L. Randall and R. Sundrum, Nucl. Phys. {\bf B557} (1999) 79, [arXiv:hep-th/9810155].}
\REF{\GAU}{D. E. Kaplan, G. D. Kribs and M. Schmaltz, Phys. Rev. {\bf D62} (2000) 035010, [arXiv:hep-ph/9911293]; Z. Chacko, M. A. Luty, A. E. Nelson and E. Ponton, JHEP {\bf 0001} (2000) 003, [arXiv:hep-ph/9911323].}
\REF{\SUM}{S. Ferrara, L. Girardello and F. Palumbo, Phys. Rev. {\bf D20} (1979) 403; P. Fayet, Phys. Lett. {\bf B84} (1979) 416;
S. Dimopoulos and H. Georgi, Nucl. Phys. {\bf B193} (1981) 150.}
\REF{\SUSY}{Y. Shadmi and Y. Shirman, Rev. Mod. Phys. {\bf 72} (2000) 25, [arXiv:hep-th/9907225]; J. Terning, [arXiv:hep-th/0306119]; K. Intriligator and N. Seiberg, Class. Quant. Grav. {\bf 24} (2007) S741, [arXiv:hep-th/0702069].}
\REF{\CHE}{M. Chemtob, Prog. Part. Nucl. Phys. {\bf 54} (2005) 71, [arXiv:hep-ph/0406029] and references therein.}
\REF{\BAR}{R. Barbier et al., Phys. Rept. {\bf 420} (2005) 1, [arXiv:hep-ph/0406039] and references therein.} 
\REF{\TAU}{R. Barate et al., Eur. Phys. Jour. {\bf C2} 91998) 395.}
\REF{\PDG}{N. Nakamatsu et al., Particle Data Group, J. Phys. {\bf G37} (2010) 075021.}
\REF{\BIS}{M. Bisset, O. C. W. Kong, C. Macesanu and L. H. Orr, Phys. Lett. {\bf B430} (1998) 274, [arXiv:hep-ph/9804282]; 
Phys. Rev. {\bf D62} (2000) 035001, [arXiv:hep-ph/9811498].}
\REF{\ROY}{S. Roy and B. Mukhopadhyaya, Phys. Rev. {\bf D55} (1997) 7020, [arXiv:hep-ph/9612447].}
\REF{\DEC}{S. Dawson, Nucl. Phys. {\bf B261} (1985) 297; H. Dreiner and P. Morawitz, Nucl. Phys. {\bf B428} (1994) 31, 
[arXiv:hep-ph/9405253].} 
\REF{\LEN}{H. Dreiner and G. G. Ross, Nucl. Phys. {\bf B365} (1991) 597.}
\REF{\COS}{K. S. Babu, T, M. Gould and I.Z. Rothstein, Phys. Lett. {\bf B321} (1994) 140, [arXiv:hep-ph/9310349]; G. Gyuk and M. Turner, Nucl. Phys. Proc. Suppl. {\bf 40} (1995) 557; A. D. Dolgov, S. Pastor, and J. W. F. Valle, Phys. Lett. {\bf B383} (1996) 193, 
[arXiv:hep-ph/9602233]; M. Kawasaki, P. Kernan, H. S. Kang, R. J. Scherrer, G. Steigman and T. P. Walker, Nucl. Phys. {\bf B419} (1994) 105; C. Liu and H. S. Song, Nucl. Phys. {\bf B545} (1999) 183, [arXiv: hep-ph/9811203].}

\singlespace
\rl{SU-ITP-11/01}
%rl{hep-ph/yymmnnn}
%\rl{\today}
%\rl{T}
\pagenumber=0
\normalspace
\medskip
\bigskip
\titlestyle{\bf{Supersymmetry Breaking by the Right--Handed Tau Neutrino}}
\smallskip
\author{ Edi Halyo{\footnote*{e--mail address: halyo@stanford.edu}}}
\smallskip
\centerline {Department of Physics} 
\centerline{Stanford University} 
\centerline {Stanford, CA 94305}
%\centerline{and}
% \centerline{California Institute for Physics and Astrophysics}
%\centerline{366 Cambridge St.}
%\centerline{Palo Alto, CA 94306}
\smallskip
\vskip 2 cm
\titlestyle{\bf ABSTRACT}
We describe supersymmetry breaking by the F-term of a heavy right-handed tau neutrino with a VEV. 
Due to the the tau neutrino Yukawa coupling, the neutralino, chargino and scalar mass matrices and the weak currents
are modified. In addition, there are new cubic and quartic scalar and trilinear R parity violating interactions. For large $\tan \beta$ these effects may be quite large. The scenario requires low energy supersymmetry breaking with generic values of 
$F \sim 10^{10}$~GeV.

\singlespace
\vskip 0.5cm
\endpage
\normalspace

\centerline{\bf 1. Introduction}
%\smallskip

According to the conventional lore, supersymmetry is broken by the nonzero F-term of a Standard Model singlet in a hidden sector. This breaking is then communicated to the observable sector, the Minimally Supersymmetric Standard Model (MSSM),
by interactions that couple to both sectors. The most popular of these mediation mechanisms are due to gravity[\GRA], gauge interactions[\OGM-\GUI], supergravity scale anomalies[\AM] and gauginos[\GAU]. The main motivation for the hidden sector is the
sum rules for the masses of particles with the same quantum numbers but different spin[\SUM]. If the supersymmetry breaking sector directly couples to the observable sector, these rules require very light superpartners that are not observed. The constraints coming from the sum rules can be avoided if supersymmetry breaking occurs in a hidden sector and is mediated to the observable sector by nonrenormalizable interactions or loop effects.
However, this does not necessarily mean that the two sectors cannot directly couple to each other but rather that the main contribution to soft supersymmetry breaking parameters should arise from loop effects or nonrenormalizable interactions and the direct couplings should be small enough not to violate the sum rules.

In this paper, we describe a scenario in which supersymmetry is broken by the nonzero F-term of the right-handed tau neutrino, $N_{\tau}=N_3$, which is assumed to be much heavier than the TeV scale. In general, all three (very heavy) right-handed neutrinos may have nonzero F-terms (and VEVs). For simplicity, here we assume that $N_e=N_1$ and $N_{\mu}=N_2$ have vanishing F-terms and VEVs.

The motivation for this scenario is twofold. First, this is an example of supersymmetry breaking which does not take place in a hidden sector. In the presence of right-handed neutrinos, $N_i$ with $i=1,2,3$, the symmetries of the MSSM allow neutrino Yukawa couplings, i.e. the superpotential contains the additional terms $y_{i\!j} L_i H_u N_j$
beyond those in the MSSM. Therefore $N_3$ directly couples to the observable sector. 
Second, if, as usual, supersymmetry is spontaneously broken the nonzero F-term of an MSSM singlet, the most economical choice is to pick a singlet that exists in the simplest extensions of the MSSM. This leads us to right-handed neutrinos which appear in many extensions of MSSM such as $SO(10)$ GUTs or numerous superstring models rather than a completely new ad hoc singlet. 

For the above reasons, we extend the MSSM by adding three (very heavy, i.e with masses $>>$TeV) right-handed neutrinos $N_i$ to the spectrum and their Yukawa couplings to the superpotential. In addition, we assume the existence of a hidden sector which breaks supersymmetry by the the right-handed tau neutrino F-term, $F_{N_3} \not= 0$ and gives rise to the VEV $N_3 \not =0$. This is not a strong assumption since there are many models of supersymmetry breaking by the nonzero F-term of a massive singlet with a VEV[\SUSY]. 

As we show below, supersymmetry breaking by the right-handed tau neutrino leads to deviations from the MSSM which are model independent, i.e. independent of the mechanism that mediates supersymmetry breaking to the observable sector. In order to stress this point, in this paper, we do not assume any specific mediation mechanism. In fact, it is highly probable that the real mediation mechanism that Nature chose is none of the ones that was considered so far in the literature. 

Since we assume that the right-handed neutrinos are much heavier than the TeV scale, they need to be integrated out
of TeV scale physics. This can be done, at the lowest order, simply by replacing the fields by their VEVs and F-terms where necessary.
By assumption, the right-handed electron and muon neutrinos do not have VEVs or F-terms and therefore integrating them out does not lead to new effects. 
{\footnote1{If $N_i$ are not much heavier than the TeV scale they also induce effective, nonrenormalizable four-fermion interactions such as $(L_i H_u)^2$ that may lead to new effects. In this paper, we neglect these.}}
However, as we show below, integrating out the right-handed tau neutrino leads to new effects since it has a VEV and F-term.

The VEV of $N_3$, spontaneously breaks R parity and lepton number leading to the effective bilinear term $L_3 H_u$ in the superpotential. This term modifies the neutralino and chargino mass matrices, changes in the neutral and charged currents and induces R parity violating trilinear interactions. In addition, there are new contributions to some scalar masses, new A and B-terms and quartic scalar interactions. These changes to the MSSM might be quite large for $\tan \beta \sim 25-45$ (together with $M_2 >280$~GeV). As a result, this scenario might be discovered at the LHC.

This paper is organized as follows. In the next section, we describe the scenario with right-handed tau neutrino supersymmetry breaking and its predictions such as changes to the neutralino and chargino mass matrices, modifications to the neutral and charged currents, new scalar mixings and interactions and induced
R parity violating trilinear terms. Section 3 includes a discussion of our results and our conclusions.

\bigskip
\centerline{\bf 2. Supersymmetry Breaking by the Right-handed Tau Neutrino}
\smallskip

We begin by adding three right-handed neutrinos, $N_i$ ($i=1,2,3$), to the MSSM spectrum.
The modified MSSM superpotential now includes the neutrino Yukawa couplings which are allowed under all the symmetries of the MSSM
$$W=y_{i\!j}^u Q_i H_u {\bar u}_j + y_{i\!j}^d Q_i H_d {\bar d}_j + y_{i\!j}^e L_i H_d {\bar e}_j + y_{i\!j}^{\nu} L_i H_u N_j + 
\mu H_u H_d, \eqno(1)$$
where $y_{i\!j}^{u,d,e,\nu}$ are $3 \times 3$ Yukawa matrices (in the flavor or generation space) for the up and down quarks and charged and neutral leptons. 
%The new neutrino Yukawa couplings are given by the term before the last one in eq. (1).
This superpotential preserves lepton number (under which $L[L_i]=1$, $L[{\bar e}_i]=L[N_i]=-1$) which we assume to be a discrete gauge symmetry{\footnote2{Discrete symmetries are necessarily gauged in string theory or in quantum gravity in which black hole effects must be taken into account.}} in order to avoid Goldstone bosons when lepton number is spontaneously broken. 
The above superpotential also preserves R parity (again assumed to be a discrete gauge symmetry) if we assign the R charge $R[N_i]=-1$
in addition to the usual R charges of the MSSM fields. We note that trilinear terms of the
form $N_i H_u H_d$ which may give rise to an unacceptably large $\mu$-term are not allowed by the lepton number symmetry. 
 
We assume that there is a hidden sector which gives rise 
to a very heavy (with mass $>>$TeV) right-handed tau neutrino with a nonzero VEV and F-term, i.e. $N_3=~<\!\!N_3\!\!> +~ \theta^2 ~F_{N_3}$. {\footnote3{With an abuse of notation, we will use $N_3$ to denote both the field and its VEV in the following.}} 
This is our main assumption which leads to all of our results. For simplicity, we assume that the other two (very heavy) 
right-handed neutrinos have vanishing VEVs and F-terms. In this paper, we are not interested in the details of the hidden sector that breaks supersymmetry since these do not affect our results. The generalization our scenario to the cases with more right-handed neutrino F-terms and VEVs is straightforward and will not be discussed in this paper. 
%However, we note that there are a large number of models in which supersymmetry is broken by the F-term of a massive scalar with a %nonzero VEV. 

The mediation of supersymmetry breaking by $F_{N_3} \not=0$ to the observable sector (by an unspecified mechanism) gives rise to soft  masses (for the squarks, sleptons, Higgs bosons and gauginos)
$$\eqalignno{
{\cal L}_m=(m_{\tilde Q})_{i\!j}^2 ~{\tilde Q_i}^* {\tilde Q_j}&+ (m_{\tilde u})_{i\!j}^2 ~{\tilde u_i}^* {\tilde u_j}+ 
(m_{\tilde d})_{i\!j}^2 ~{\tilde d_i}^* {\tilde d_j}+ (m_{\tilde L})_{i\!j}^2 ~{\tilde L_i}^* {\tilde L_j} & (2) \cr 
&+(m_{\tilde e})_{i\!j}^2 ~{\tilde e_i}^* {\tilde e_j}
+(m_{\tilde \nu})_{i\!j}^2 ~{\tilde \nu_i}^* {\tilde \nu_j}+ m^2_H (H_u^2+H_d^2)+ M_i {\bar \lambda_i} {\lambda_i}, \cr  
}$$
where we did not include the new soft masses squared for the right-handed sneutrinos since they are very heavy (due to their supersymmetric masses). 
In addition, the scalar potential contains the A (trilinear scalar couplings) and B-terms (bilinear scalar couplings)
$${\cal L}_{A,B}=A_{i\!j}^u y_u {\tilde Q}_i H_u {\tilde u}^*_j + A_{i\!j}^d y_d {\tilde Q}_i H_d {\tilde d}^*_j + A_{i\!j}^e y_e {\tilde L}_i H_d {\tilde e}^*_j + A_{i\!j}^{\nu} y_{\nu} {\tilde L}_i H_u N_j + B_1 \mu H_u H_d. \eqno(3)$$
Explicit expressions for the soft masses and the coefficients $A$ and $B$ can be obtained only in the context of a specific mediation mechanism. We assume that the squark and slepton mass squared matrices are such that flavor-changing neutral currents are suppressed as they are for example in gauge or anomaly mediation.

In addition, for simplicity, we assume that the neutrino Yukawa coupling matrix, $y_{i\!j}^{\nu}$, in eq. (1) has only one nonzero entry, $y_{33}^{\nu}=y \not =0$. (More generally we can take also $y_{1i}, y_{2i} \not=0$  since we are setting the VEVs and F-terms of the other two right-handed neutrinos $N_1,N_2$ to zero.) That is, $N_3$ has a Yukawa coupling only to $\nu_{\tau}$ and not to $\nu_e$ or $\nu_{\mu}$. 
Thus, the model introduces three new parameters (in addition to those in the MSSM) which are the tau neutrino Yukawa coupling $y$, the VEV $N_3$ and the F-term, $F_{N_3}$. We assume that these parameters are real in order not to introduce new CP violation phases. 

Since the right-handed neutrinos are very heavy we can integrate them out, at the classical level, by replacing the fields $N_i$ by their VEVs and F-terms where necessary. When $N_3 \not=0$, lepton number and R parity are spontaneously broken and
$$W_{N}=y L_3 H_u N_3=yN_3(\nu_{\tau} H_u^0 - \tau_L H_u^+) \eqno(4)$$
effectively generates the lepton number and R parity violating bilinear term $\epsilon L_3 H_u$ with coefficient $\epsilon=yN_3$. This leads to a number of results that have been studied in the context of R parity violation[\CHE,\BAR].
%{\footnote4{This includes LSPs which become unstable and therefore cannot be the dark matter.}}
Most (but not all) of our results follow directly from this observation. Below we list the most important predictions of supersymmetry breaking by the right-handed tau neutrino.

\noindent
{\it 2.1. The Neutralino Mass Matrix}~:
When $N_3 \not=0$, we see from eq. (4) that $\nu_{\tau}$ mixes with $H_u^0$ or the
neutralinos. Therefore, the neutralino mass matrix becomes $5 \times 5$ and is given by (in the basis ${\tilde B}, {\tilde W}, 
{\tilde H_u^0}, {\tilde H_d^0}, \nu_{\tau}$)
$$M_{{\tilde \chi}^0}=\pmatrix{
  M_1 & 0 &  g^{\prime} v_u/2  &  -g^{\prime} v_d/2  & 0 \cr
   0 & M_2  &  -g v_u/2  &  g v_d/2  &  0  \cr
   g^{\prime} v_u/2  &  -g v_u/2  &  0  & -\mu  & -yN_3  \cr
   -g^{\prime} v_d/2  &  g v_d/2  &  -\mu  &  0  &  0  \cr
   0  &  0  &  -yN_3  &  0  &  0  \cr
} \eqno(5)$$
where $M_1,M_2$ ($g_1,g_2$) are the $U(1)_Y$ and $SU(2)_L$ gaugino masses (gauge couplings) respectively, and $v_u,v_d$ are the 
VEVs of $H_u^0,H_d^0$.
The neutralino mass matrix is modified from its MSSM form by the new mixing terms given by $yN_3$. 
A strong constraint on the magnitude of $yN_3$ arises from the bound on the Majorana tau neutrino mass. In this paper, we use the bound from accelerator experiments, i.e. $m_{\nu_\tau}< 18.2$~MeV[\TAU,\PDG]. The cosmological bound on stable neutrino masses[\PDG],
$\Sigma m_{\nu}< 1$~eV, is much stronger but can be evaded if an MeV scale tau neutrino can decay. This is expected to be the case when R parity is violated as we discuss in section 2.7.

One can obtain a bound on $y N_3$ in terms of the tau neutrino mass bound[\BIS]
$$(yN_3)^2<{{4x \mu^2  m_{\nu_\tau}} \over {v^2 \cos^2 \beta (xg^2+ g^{\prime 2})- 4x M_2  m_{\nu_\tau}}}\quad, \eqno(6)$$
where $v^2=v_u^2+v_d^2=(174)^2$~GeV$^2$, $\tan \beta=v_u/v_d$ and $x=5 \tan^2 \theta_W/3$.
It is easy to see that, for $\tan \beta \sim 45$ the bound on $yN_3$ is quite large, $yN_3< 500$~GeV (depending on $\mu^2/v^2$ which we assume to be $\sim 1$).
If in addition, $M_2 \sim 280$~GeV there is no upper bound on $yN_3$ since the denominator in eq. (6) vanishes. Clearly a lower $\tan \beta$ may also lead to an unbound $yN_3$. For example, if $\tan \beta \sim 25$, an unbound $yN_3$ requires $M_2 \sim 1$~TeV which we consider to be the upper limit due to naturalness. As a result, $yN_3$ can be as large as the TeV scale and the corrections to the neutralino (chargino and scalar) masses may be very large. However, the tau neutrino mass is only one of many constraints on 
the magnitude of $y N_3$ and it is not clear whether such a large value satisfies the other constraints.

\noindent
{\it 2.2. The Chargino Mass Matrix}~:
Eq. (4) also shows that $\tau_L$ mixes with $H_u^+$ or the charginos. Therefore the chargino mass matrix becomes $3 \times 3$ and is given by (in the basis ${\tilde W^+}, {\tilde H_u^+}, {\bar \tau}_R$ and their charge conjugates ${\tilde W^-}, {\tilde H_u^-}, 
{\tau_L}$)
$$M_{{\tilde \chi}^{\pm}}=\pmatrix{
    M_2  &  g v_u/2  &  0  \cr
    g v_d/2  &  \mu  &  0  \cr
    0  &  yN_3  &  m_3  \cr
}  \eqno(7)$$ 
A subtle but important issue is the identity of $\tau_{L,R}$ that appear in the chargino basis. These $\tau_{L,R}$ are the eigenstates of the charged lepton mass matrix, $y_{i\!j}^e v_d$, and therefore a linear combination of the interaction eigenstates
$e, \mu, \tau$ that appear in eq. (1). This is crucial for the decay of (the massive) $\nu_{\tau}$ which is required to evade the cosmological bound on neutrino masses.

Again, the deviations of the chargino mass matrix from its MSSM form are parametrized by $yN_3$ and can be quite large.
Diagonalizing $M_{\tilde \chi^{\pm}} M_{\tilde \chi^{\pm}}^{\dagger}$ and demanding $m_{\tau} \simeq m_3$ gives the approximate analytic formulas for chargino masses (assuming $yN_3>>m_3$)[\BIS]
$$\eqalignno{
M^2_{{\tilde \chi^+}\!,{\tilde \chi^-}}=&{1 \over 2}[M_2^2+2M_W^2+ \mu^2+(yN_3)^2] & (8)  \cr
&\pm{1 \over 2}[(\mu^2+(yN_3)^2-M_2^2-2M_W^2 \cos 2 \beta)^2+8M_W^2(M_2 \sin \beta+ \mu \cos \beta)^2]^{1/2}.  \cr
}$$
We note that $yN_3$ enters the mass formulas in way very similar to $\mu$ and chargino masses reduce to their MSSM values for
$N_3=0$. When $N_3 \not =0$ the smaller chargino mass increases relative to its MSSM value.

\noindent
{\it 2.3. Neutral and Charged Currents}~:
In order to obtain the physical mass eigenstates, the neutralino and chargino mass matrices have to be diagonalized, which
means that the mass eigenstates are linear superpositions of the interaction eigenstates. Thus, diagonalization of
$M_{{\tilde \chi}^0}$  gives rise to a $5 \times 5$ neutralino mixing matrix N, and that of $M_{{\tilde \chi}^{\pm}}$ to $3 \times 3$ left and
right-handed chargino mixing matrices U and V. These mixings lead to deviations of the neutral and charged currents from their MSSM form. The neutral currents are now given by[\ROY]
$${\cal L}_{{\tilde \chi}^+ {\tilde \chi}^- Z^0}={g_2 \over {2 \cos \theta_W}} {\tilde \chi}_a^+ \gamma^{\mu}(P_L  A_{ab}^L+
P_R A_{ab}^R) {\tilde \chi}_b^-  Z_{\mu}^0+h.c. , \eqno(9)$$
$${\cal L}_{{\tilde \chi}^0 {\tilde \chi}^0 Z^0}={g_2 \over {2 \cos \theta_W}} {\tilde \chi}_a^0 \gamma^{\mu}(P_L B_{ab}^L+
P_R B_{ab}^R) {\tilde \chi}_b^0  Z_{\mu}^0 , \eqno(10)$$
whereas the charged currents are given by
$${\cal L}_{{\tilde \chi}^+ {\tilde \chi}^0 W^-}={g_2 \over {\sqrt{2}}} {\tilde \chi}_a^+ \gamma^{\mu}(P_L C_{ab}^L+
P_R C_{ab}^R) {\tilde \chi}_b^0  W_{\mu}^+ +h.c.   \eqno(11)$$
Above, ${\tilde \chi}_a^{\pm}$ and ${\tilde \chi}_a^0$ are the chargino and neutralino mass eigenstates respectively and
$P_{R,L}={1 \over 2}(1 \pm \gamma_5)$. The matrices $A^{L,R} ,B^{L,R}, C^{L,R}$ are determined by the
mixing matrices N,U,V[\ROY].
 
As a result, the gauge couplings are no longer universal, i.e. those of the third generation leptons are different from those of the first two generations. In addition, gauge interactions mix $\tau$ ($\nu_{\tau}$) with the charginos (neutralinos), i.e. the gauge interactions are not diagonal. For example, there are new vertices such as $Z^0 {\tilde \chi}_i^{\pm} e_j^{\mp}$, $Z^0 {\tilde \chi}_i^0 \nu_j$, $W^{\pm} {\tilde \chi}_i^{\mp} \nu_j$ and $W^{\pm} {\tilde \chi}_i^0 e_j^{\mp}$.  Among these vertices,
the ones that do not contain the third generation leptons are suppressed relative to the others due to the extra charged lepton mixing required for them.

\noindent
{\it 2.4. Induced Trilinear R Parity Violating Operators}~:
Since we assume that R parity and lepton number are discrete gauge symmetries of the model, the original superpotential in eq. (1) does not include the R parity and lepton number ($\Delta L=1$) violating terms
$$W_L=\lambda_{i\!jk} L_i L_j {\bar e}_k+ \lambda_{i\!jk}^{\prime} L_i Q_j {\bar d}_k,  \eqno(12)$$
$W_L$ above gives rise to the interactions
$${\cal L}_{\lambda}=\lambda_{i\!jk}[{\tilde \nu}^i_L e_L^j {\bar e}^k_R+ {\tilde e}^j_L \nu_L^i {\bar e}^k_R+{\tilde e}^{k*}_R 
e_L^j {\bar \nu}^i_L-(i \leftrightarrow j)]+ h.c.,  \eqno(13)$$
$${\cal L}_{\lambda^{\prime}}=\lambda^{\prime}_{i\!jk}[{\tilde \nu}^i_L d_L^j {\bar d}^k_R+{\tilde d}^j_L \nu_L^i {\bar d}^k_R+
{\tilde d}^{k*}_R d_L^j {\bar \nu}^i_L- {\tilde e}^i_L u_L^j {\bar d}^k_R- {\tilde u}^j_L e_L^i {\bar d}^k_R-
{\tilde d}^{k*}_R u_L^j {\bar e}^i_L]+h.c. \eqno(14)$$
There are 9 $\lambda_{i\!jk}$ due to the antisymmetry of the first two indices and 27 $\lambda_{i\!jk}^{\prime}$ terms.

However, the mixing of $\tau_L$ with charginos and of $\nu_{\tau}$ with neutralinos, combined with the gaugino and Higgsino interactions, induce some of the above $\lambda$ and $\lambda^{\prime}$ terms[\ROY]. For example, due to the mixing, 
the gaugino-lepton-slepton coupling $g {\tilde \nu}^i_L e_L^i {\tilde W}^+$ gives rise to the interaction $g V_{31}^* {\tilde \nu}^i_L e_L^i {\bar \tau}_R$ which is of the type in eq. (13) where $\lambda_{ii3}=g V_{31}^*$ and 
$V_{31}$ is an element of the left-handed chargino mixing matrix $V$. Similarly, Higgsino couplings to the
third generation leptons and sleptons (especially for large $\tan \beta$) generate $\lambda$ and $\lambda^{\prime}$ terms. For example, the interaction
$y_{\tau} {\tilde \tau}_L {\bar \tau}_R {\tilde H}_d$ gives rise to the $\lambda_{333}$ term $y_{\tau} N_{35} {\tilde \tau}_L 
{\bar \tau}_R  \nu_{\tau L}$ where $N_{35}$ is an element of the neutralino mixing matrix $N$. The $\lambda^{\prime}$ terms are obtained in a similar fashion from the gaugino and (third generation) Higgsino couplings to the quarks and squarks. Note that in
general, a specific $\lambda_{i\!jk}$ or $\lambda_{i\!jk}^{\prime}$ term gets multiple contributions from different mixings and/or
gaugino and Higgsino interactions. 

%The Yukawa couplings of the first two generations of leptons are too small to generate trilinear terms with appreciable coefficients
%and we neglect them below. (See however, section 2.7 where these might be crucial for the decay of the massive tau neutrino.)
There are a large number of $\lambda_{i\!jk}$ and $\lambda^{\prime}_{i\!jk}$ terms that are induced from the gaugino and Higgsino interactions. We do not list them here but discuss a number of important properties they exhibit. 
%For simplicity and because they are very small, below we neglect the Yukawa couplings of the first two generations of leptons. 
\smallskip
\parskip=0pt
\item{1)} The induced terms have an interesting index structure if we neglect charged lepton/slepton mixing[\ROY] (and there is only one nonzero VEV $N_3$). If the mixing is taken into account, all $\lambda_{i\!jk}$ and $\lambda_{i\!jk}^{\prime}$ are induced but the ones not mentioned below are suppressed. 
The induced $\lambda_{i\!jk}$ have one index equal to $3$ (since only the third generation leptons mix
with neutralinos and charginos) and the other two indices equal to each other. 
Thus, only 4 of the 9 $\lambda_{i\!jk}$ in eq. (13) are induced. For the $\lambda^{\prime}_{i\!jk}$, only terms with $i=3$ are induced for the same reason above. Thus, only 9 of the 27 $\lambda^{\prime}_{i\!jk}$ in eq. (14) are induced.
\item{2)} There are new induced $\lambda$ and $\lambda^{\prime}$ terms that do not appear in eqs. (13) and (14) since they are not invariant under $SU(2)_L$[\ROY]. These terms have ``wrong" helicity structures such as $RRL$ or $LLL$ which are not gauge invariant. For example, the gaugino interaction $g {\tilde e}_L^i e_L^i {\tilde B}^0$ gives rise to the term $g N_{15} {\tilde e}_L^i e_L^i \nu_{\tau L}$ with the $LLL$ structure where $N_{15}$ is an element of the neutralino mixing matrix. 
\item{3)} From eq. (13) it is clear that $\lambda_{i\!jk}$ are antisymmetric in the first two indices $i,j$. This is not the case for 
some of the induced terms. For example, the term we considered above (induced by the gaugino interactions), $g V_{31}^* {\tilde \nu}^i_L e_L^i {\bar \tau}_R$, has the first two indices identical and therefore symmetric (again in violation of $SU(2)_L$). These
terms and the ones with the ``wrong" helicity structure give rise to new lepton number violating operators that are absent in the conventional R parity violating version of the MSSM.
\item{4)} The coefficients of all induced lepton number violating operators are calculable[\ROY]. 
For given values of the parameters $yN_3$ and $yF_N$ (in addition to the MSSM parameters), we can diagonalize all the
fermion mass matrices and using the mixing matrices U,V,N, obtain all the coefficients of the induced $\lambda$ and $\lambda^{\prime}$ terms. 
\item{5)} Operators that violate baryon number such as ${\bar u}^i_R {\bar d}^j_R {\bar d}^k_R$ are not induced. This is fortituous since the bounds on the coefficients of baryon number violating operators are extremely strong.  
%\item{6)} Even though lepton number is spontaneously broken, neutrinoless double beta decays are not predicted since the physics of %the first two generation of leptons has not been modified.
\parskip=5pt

\noindent
{\it 2.5. New Scalar Interactions}~:
In the scalar sector, the Higgs doublets $H_u=(H_u^+, H_u^0)$ and $H_d=(H_d^0, H_d^-)$ now mix with
the slepton doublet ${\tilde L}=(\tilde {\nu_{\tau}}, {\tilde \tau})_L$ and ${\tilde \tau}_R$.
These 14 real fields are divided into three neutral scalars, three pseudoscalars and four charged scalars (and their conjugates). Out of these, one neutral and two charged scalars are eaten by the gauge bosons through the Higgs mechanism
leaving 11 real fields: two neutral scalars, three pseudoscalars and three charged scalars. 

The complete scalar potential arises from the F-terms, A and B-terms and the
$SU(2)_L$ and $U(1)_Y$ D-terms. We do not explicitly write it here since it is too cumbersome[\ROY] and our only aim is to highlight the differences between the scalar potential in our scenario and that of the MSSM. 
%We do not explicitly write the scalar mass squared matrices which appear in [\ROY]. 
As we discuss below, in our scenario, there are new contributions to the diagonal and off-diagonal elements of the scalar mass matrices that modify scalar masses and the conditions for acceptable electro-weak symmetry breaking. In addition, there are new
trilinear and quadratic scalar couplings that are absent in the MSSM.

Due to the neutrino Yukawa term in the superpotential, there are new contributions to the scalar potential. The new terms coming from 
$|F_{N_3}|^2$, $|F_{L_3}|^2$ and $|F_{H_u}|^2$ give rise to
$$\eqalignno{
V_s=y^2 |N_3|^2(|{\tilde \nu_{\tau}}|^2&+|H_u^0|^2+|{\tilde \tau}_L|^2+|H_u^+|^2)+ y^2(|{\tilde \nu_{\tau}} H_u^0|^2+
|{\tilde \tau}_L H_u^+|^2  &  (15) \cr
&- {\tilde \nu_{\tau}} H_u^0 {\tilde \tau}_L H_u^+ +h.c.)  +yF_{N_3} ({\tilde \nu_{\tau}} H_u^0-{\tilde \tau}_L H_u^+)+h.c.). \cr
}$$
The first term in eq. (15) gives new contributions of $y^2 N_3^2$ to the scalar masses squared $m^2_{H_u},m^2_{\tilde {\tau}_L},m^2_{\tilde {\nu_{\tau}}}$ beyond the soft masses. The second term describes new quartic scalar interactions 
of strength $y^2$ which do not appear in the MSSM.
The third term arises from $F_{N_3} \not=0$ and gives rise to new mixings between either ${\tilde {\nu_\tau}}$ and $H_u^0$ or ${\tilde \tau}_L$ and $H_u^+$.
This contributes to the B-term for the effective bilinear term $B_2 \epsilon L_3 H_u$ (where $\epsilon=yN_3$) with 
$B_2=F_{N_3}/N_3$. There is also the A-term arising from the tau neutrino Yukawa coupling, $A y N_3 L_3 H_u$ which gives another contribution $B_2=A$ so that overall we have the effective B-term $B_2= A+F_{N_3}/N_3$.

In addition, there are new contributions to the scalar potential coming from the mixed terms. These  
new bilinear and trilinear scalar interactions are given by
$$V_s= y \mu N_3 ({\tilde L_3} H_d^{\dagger}+ h.c.)+ y y_{i\!j}^u N_3 ({\tilde Q_i} {\tilde u_j}^* {\tilde L_3}^{\dagger} + h.c.) 
+ y y_{\tau} N_3 (H_u^{\dagger} H_d {\tilde \tau}_R^* + h.c.). \eqno(16)$$
The first term is another slepton-Higgs mixing term that contributes to the off-diagonal elements of scalar mass squared matrices.
This can be seen as a new B-term without a corresponding term in the superpotential in eq. (1). 
The last two terms are completely new trilinear scalar interactions (A-terms) which do not arise from terms in the superpotential, i.e. from those in eq. (1) or the trilinear R parity violating terms in eq. (12). We note that these new A-terms have coefficients that are not free but fixed to be $(y_u,y_{\tau})~ y N_3$.

\noindent
{\it 2.6. The Lightest Supersymmetric Particle (LSP)}~:
In the R parity preserving MSSM, the LSP has to be charge and color neutral for cosmological reasons and therefore is expected to be the lightest neutralino, ${\tilde \chi}_1^0$. On the other hand, when R parity is broken the LSP might be a neutralino, chargino, squark or slepton. Moreover, due to the spontaneous R parity breaking (or the explicit breaking by the induced $\epsilon L_3 H_u$ term), the LSP is no longer stable. If the LSP is still a neutralino and $m_{{\tilde \chi}_1^0}>M_Z$, then it decays into 
$\tau^{\pm} W^{\mp}$ and $Z^0 \nu_{\tau}$ due to gauge interactions whereas if $m_{{\tilde \chi}_1^0}<M_Z$ its decays are into 
$\tau f {\bar f}^{\prime}$ and $\nu_{\tau} f {\bar f}$ (where $f$ is a quark or lepton) due to $\lambda$ and $\lambda^{\prime}$ interactions. For example, a light
neutralino LSP that is mainly ${\tilde \gamma}$ with a dominant coupling $\lambda_{131}^{\prime}$ has a decay rate[\DEC]
$$\Gamma_{LSP}={{3 \alpha_{em} \lambda_{131}^{\prime~ 2}} \over {128 \pi^2}} {M_{LSP}^5 \over m_{\tilde f}^4}\quad, \eqno(17)$$
where $m_{\tilde f}$ is the mass of the intermediate sfermion in the decay.

Whether the LSP decays inside the detector depends on its mass and the strength of the its interactions. For example, for the light neutralino LSP above, the decay length (in meters) is given by[\LEN] 
$$L=3 \times 10^{-3} (\beta \gamma) \left({m_{\tilde f} \over 100~GeV} \right)^4 \left({1~GeV \over M_{LSP}} \right)^5
\left(1 \over {\lambda_{131}^{\prime~ 2}} \right), \eqno(18)$$
where $\gamma$ is the Lorentz factor.
If ${\tilde \chi}_1^0$ does not decay inside the detector, its behavior in the lab is identical to a stable LSP. However, for all
phenomenologically relevant values e.g. for $\lambda, \lambda^{\prime}>10^{-20}$ the 
neutralino LSP decays during the lifetime of the universe and can no longer be the dark matter.

\noindent
{\it 2.7. Neutrino Masses}~:
As we saw in section 2.1, in order to get $yN_3 \sim 500$~GeV, it is crucial to use the lab bound $m_{\nu_\tau}<18.2$~MeV[\TAU,\PDG] in eq. (6). However, the cosmological bound on the neutrino masses, $\Sigma m_{\nu}<1$~eV[\PDG], is much stronger. This bound can only be evaded if $\nu_{\tau}$ is not stable[\COS] which is precisely the situation when R parity is broken.
For example, due to the induced trilinear R parity (and lepton number) violating interactions $\nu_{\tau}$ can decay through $\nu_{\tau} \to {\tilde \chi}_1^{0*} {\tilde \nu}_{\tau}^*$ followed by (for small masses) ${\tilde \chi}_1^{0*} \to W^{\pm*} e^{\mp} \to e^- e^+ \nu_e$ and ${\tilde \nu}_{\tau}^* \to e^- e^+$. Thus, the overall decay is $\nu_{\tau} \to 2e^-+2e^+ +\nu_e$. In our scenario, the above ${\tilde \chi}_1^0$ decay is only possible due to charged slepton (${\tilde \tau}$-${\tilde e}$) mixing. The source of this mixing is either the diagonalization of the charged lepton mass matrix, $y_{ij}^e v_d$ in eq. (1) or the slepton soft mass matrix in eq. (2) which in general is not diagonal. 
In addition, there may be other $\nu_{\tau}$ decay channels depending on $m_{{\tilde \chi}_1^0}$ and
the complete list of induced $\lambda$ and $\lambda^{\prime}$ interactions. Whether the cosmological bound is evaded due to the decay of tau neutrinos depends on the magnitude of the induced trilinear couplings and the amount of charged lepton and slepton mixings. Since all that is required is a $\nu_{\tau}$ lifetime which is less than the that of the universe, even very small $\lambda$ and $\lambda^{\prime}$ couplings are sufficient for our purposes.

When R parity is broken, neutrino masses are also generated radiatively by one-loop diagrams involving the $\lambda$ and $\lambda^{\prime}$ interactions[\CHE, \BAR]. The strongest bound arises from the $\nu_e$ mass, $m_{\nu_e}<2.2~eV$[\PDG]. This, in turn, gives rise to the bounds $|\lambda_{133}|<9.4 \times 10^{-4}$ and $|\lambda_{133}^{\prime}|<2.1 \times 10^{-4}$ (for $m_{\nu}=.30$~eV and sparticle masses of $\sim 100$~GeV)[\BAR]. 
In our scenario, these strong bounds can be satisfied since $\lambda_{133}$ and $\lambda_{133}^{\prime}$ are suppressed due to the very small charged lepton and/or slepton mixings that are required. We conclude that one-loop neutrino masses do not pose a problem for our scenario.

\noindent
{\it 2.8. Parameters and Scales}~:
If we make no assumptions about the mediation mechanism of supersymmetry breaking, the neutralino and chargino mass matrices contain the six parameters 
$M_1,M_2, \tan \beta, \mu, yN_3, m_3$. If we know all the eight neutralino and chargino masses we can determine these six parameters
(In fact there have to be two relations among them.). Note that the only way to determine $y$ is by measuring the new quartic scalar couplings since that is the only place $y$ appears on its own. In order to determine $yF_{N_3}$ we need information about the scalar masses since $yF_{N_3}$ appears only in the scalar mixing terms. 

The new parameters $y$, $yN_3$ and $yF_{N_3}$ are somewhat constrained even without any assumptions on supersymmetry breaking mediation mechanism. In order to avoid a large Dirac tau neutrino mass, $y$ has to be quite small, $y<10^{-4}$. Such a small Yukawa coupling is not natural in the 't Hooft sense, i.e. the symmetry of the
model does not increase when the coupling is set to zero. On the other hand, we know nothing about the physics that determines the Yukawa couplings. Moreover, the electron Yukawa coupling certainly shows that these can be very small, e.g. $y_e \sim 10^{-5}$. 
Thus, $y \sim 10^{-4}$ is not such an unreasonable assumption.
For TeV scale soft masses with no fine-tuning, we need $y N_3<10^3$~GeV and $y F_{N_3}<10^6$~GeV$^2$ which in turn give $N_3<10^7$~GeV and $F_{N_3}<10^{10}$~GeV$^2$. We see that the effective supersymmetry breaking scale has to be relatively low which can be easily accomodated in gauge and anomaly mediation mechanisms. However, it is very hard to naturally realize this in gravity mediation due to the much larger supersymmetry breaking scale in that scenario. Of course, by taking a much smaller value for $y$ we can increase the scales of $N_3$ and $F_{N_3}$ but this makes the model less realistic.
%y \sim 10^{-15} in gravity mediation

%how to distinguish between scenarios
\noindent
{\it 2.9. Signatures of the Scenario}~:
Consider a different scenario with just the bilinear term $\epsilon L_3 H_u$ added to the
MSSM superpotential (with no right-handed neutrinos and supersymmetry breaking by another singlet). Most of the signatures of this  scenario are identical to the ones discussed above. On the other hand, in this case, there are no new scalar interactions, i.e. the A-terms and the quartic interactions in eqs. (15) and (16) are absent. (The extra contributions to the scalar masses squared are also absent but it is very hard to disentangle these from the regular soft masses.) Since $y<10^{-4}$ the quartic interactions are probably very hard to detect. However, the two scenarios differ by the new A-terms, i.e. the new trilinear scalar interactions which are quite large especially for the third generation squarks and sleptons with large Yukawa couplings.

Another possibility is a scenario in which right-handed tau neutrinos get VEVs and effectively generate the $\epsilon L_3 H_u$ term but supersymmetry breaking is due to another singlet. This case reproduces all our results except for the corrections to the scalar mixings which arise from $F_{N_3}$. Thus, in general, distinguishing between these two scenarios requires the knowledge of scalar mixings in detail. 
This task becomes easier in the context of a specific mediation mechanism which determines the soft masses and A and B-terms. For example, in gauge mediation, all squark, slepton and gaugino masses are given in terms of one variable, $\Lambda=F_{N_3}/N_3$. In addition, all A-terms vanish (or are very suppressed). $\Lambda$ which
can be easily fixed by, for example, the squark masses is exactly the ratio between the two parameters introduced in our scenario, i.e. $\Lambda=y F_{N_3}/ yN_3$. As we argued above, $yN_3$ and $yF_{N_3}$ can be determined separately from the physical sfermion and scalar masses respectively. Their ratio tells us whether supersymmetry is broken by the F-term of the right-handed tau neutrino (assuming gauge mediation). Alternatively, in this case we find $B_2=\Lambda$ so determining the scalar mixing terms in eq. (15) is crucial for experimentally verifying our scenario. 
 
Finally, the variation of our scenario in which supersymmetry is broken by the right-handed tau neutrino with a vanishing VEV is very hard to distinguish from one in which supersymmetry is broken by another singlet. Since $N_3=0$ in this case, R parity and lepton number are not spontaneously broken and therefore neutralino and chargino mass matrices do not receive corrections, neutral and charged currents are not modified and there are no new A-terms. The only new effects are the new B-terms, i.e. the scalar mixing terms mass given by eq. (15) which may be large and determined from the physical scalar masses.

%It is not clear whether these two effects are large enough to experimentally distinguish between supersymmetry breaking by the right-%handed tau neutrino and another MSSM singlet.

\bigskip
\centerline{\bf 3. Conclusions and Discussion}
\smallskip

% repeat, scenario with predictions and experimentally verifiable, tau neutrino mass bound
In this paper, we introduced a new scenario in which supersymmetry is broken by the F-term of a heavy right-handed tau neutrino with
a nonzero VEV. Supersymmetry breaking may be communicated to the observable sector by any mediation mechanism in addition to the tau neutrino Yukawa coupling. Due to the latter, supersymmetry is not broken in a hidden sector. The scenario has a number of predictions which arise from the VEV of $N_3$, leading to the effective (R parity and lepton number violating) bilinear term in the superpotential, e.g. corrections to the neutralino and chargino mass matrices, changes in the neutral and charged currents and induced R parity violating interactions. However, there are also predictions beyond those arising from the effective bilinear term, e.g. extra contributions to scalar masses, new A and B-terms and quartic scalar interactions. These new effects might be quite large for $\tan \beta \sim 25-45$ (together with $M_2 >280$~GeV). As a result, this scenario might be experimentally verified at the LHC.
%bound on tau neutrino mass and tan beta

%things to do, parameter space scan taking into the bounds, compute all lambdas, compute neutral and charged currents etc
Clearly, in this paper, we did not thoroughly examine the implications of supersymmetry breaking by the right-handed tau neutrino. Much more work needs to be done in order to understand the viability of this scenario. First, the parameter space of the scenario (together with that of the MSSM) should be scanned taking into account as many experimental bounds as possible. Second,
all corrections to neutral and charged currents that arise from the new mixings should be computed as functions of the parameters. Third, coefficients of all induced R parity violating trilinear terms, including those that are new and unconventional, should be computed at different points in the parameter space. 
The corrections to the weak currents and the trilinear R parity violating interactions are expected to give rise to many new processes that may either bound the parameters or experimentally verify (or falsify) our scenario.

%experimetal verification, Z0 leptonic decays universality, LR asymmetry, lepton hadron flavor violating decays from lambdas
Assuming that we know all the corrections to the neutral and charged currents together with all the induced $\lambda$ and $\lambda^{\prime}$ terms, there are a number of processes that may verify our scenario. For example, due to the corrections to the neutral currents,
the $Z^0$ couplings to leptons are no longer universal. The $Z^0$ decay rate into taus is different from those into electrons and muons. Moreover, these corrections are not left-right symmetric; as a result, the left-right asymmetry in leptonic $Z^0$ decays is different for taus than for the other charged leptons. In addition, the induced R parity violating trilinear terms give rise to lepton and hadron flavor violating decays that might be detectable in $B$-${\bar B}$ mixing, $\Delta m_K$ and leptonic rare $B$ decays.
%LHC physics such as single slepton production, particle decays, LSP decays,completely new effects from A terms quartic couplings
Finally, due to R parity violation, LHC phenomenology is modified. Single production of sleptons becomes possible even though the
dominant production mechanism remains pair production. In addition, supersymmetric particles may decay directly into particles
through the induced $\lambda$ and $\lambda^{\prime}$ interactions even though cascade decays to the LSP through gauge interactions are dominant. In particular, as we mentioned in section 2.6, the LSP becomes unstable with an interesting phenomenology.

%three N_i case similar to GGM
In this paper, for simplicity, we assumed that only $N_3 \not=0$ and $F_{N_3} \not =0$, i.e. supersymmetry is broken by only the right-handed tau neutrino. Clearly, this can be easily generalized to the case with F-terms and VEVs for (two or) the three 
right-handed neutrinos. 
In this case, all three charged leptons mix with the charginos resulting in a $5 \times 5$ chargino mass matrix and all three left-handed neutrinos mix with the neutralinos giving rise to a $7 \times 7$ neutralino mass matrix. 
There are no strong bounds on the new F-terms, $F_{N_1}$ and $F_{N_2}$, and they can be as large as $F_{N_3}$. 
However, there are very (relatively) strong constraints on the VEV of $N_e=N_1$ ($N_{\mu}=N_2$). For example, neutrinoless double beta decays give rise to the bound $y_1N_1<10^{-4}~yN_3$. Even for $\tan \beta \sim 45$ for which $yN_3$ can be in the hundreds of GeVs, we find $y_1N_1 <10^{-1}$~GeV which is very small. The decays of a massive neutrino by a virtual $W$ boson gives the bound $y_2N_2< 1$~GeV for $\tan \beta \sim 45$ which is also quite small. These small values for $y_1N_1$ and $y_2N_2$ might be due to small VEVs,
e.g. $N_1 \sim 1$~GeV and $N_2 \sim 10$~GeV or alternatively due to very small Yukawa couplings, e.g. $y_1 \sim 10^{-7}$ and
$y_2 \sim 10^{-6}$. It would be interesting to generalize our scenario to the case with three nonzero $F_{N_i}$ and $N_i$ and find out its generic predictions.

%other visible sector sysy breaking, NMSSM
%how to tie up at the end

\bigskip
\centerline{\bf Acknowledgements}
\noindent
I would like to thank the Stanford Institute for Theoretical Physics for hospitality.

\vfill

\refout

\end
\bye